\documentclass[twocolumn]{aastex63}

\usepackage{csquotes}
\usepackage{gensymb}
\usepackage{url}

\newcommand{\mycenter}{\centering\leftskip=0pt minus0.5in \rightskip=\leftskip}

\received{10 October 2022}
\revised{22 February 2023}
\accepted{23 February 2023}
\submitjournal{ApJ}

\graphicspath{{./}{figures/}}

\begin{document}

\title{First BISTRO observations of the dark cloud Taurus L1495A-B10:
the role of the magnetic field in the earliest stages of low-mass star formation}

\author[0000-0003-1140-2761]{Derek Ward-Thompson}
\affiliation{Jeremiah Horrocks Institute, University of Central Lancashire, Preston PR1 2HE, UK}

\author[0000-0001-5996-3600]{Janik Karoly}
\affiliation{Jeremiah Horrocks Institute, University of Central Lancashire, Preston PR1 2HE, UK}

\author[0000-0002-8557-3582]{Kate Pattle}
\affiliation{Department of Physics and Astronomy, University College London, WC1E 6BT London, UK}

\author[0000-0002-1178-5486]{Anthony Whitworth}
\affiliation{School of Physics and Astronomy, Cardiff University, The Parade, Cardiff, CF24 3AA, UK}

\author[0000-0002-4552-7477]{Jason Kirk}
\affiliation{Jeremiah Horrocks Institute, University of Central Lancashire, Preston PR1 2HE, UK}

\author[0000-0001-6524-2447]{David Berry}
\affiliation{East Asian Observatory, 660 N. A{'}oh\={o}k\={u} Place, University Park, Hilo, HI 96720, USA}

\author[0000-0002-0794-3859]{Pierre Bastien}
\affiliation{Centre de recherche en astrophysique du Qu\'{e}bec \& d\'{e}partement de physique, Universit\'{e} de Montr\'{e}al, C.P. 6128 Succ. Centre-ville, Montr\'{e}al, QC, H3C 3J7, Canada}

\author[0000-0001-8516-2532]{Tao-Chung Ching}
\affiliation{Zhejiang Lab, Kechuang Avenue, Yuhang District, Hangzhou 311121, People{'}s Republic of China}

\author[0000-0002-0859-0805]{Simon Coud\'{e}}
\affiliation{Department of Earth, Environment, and Physics, Worcester State University, Worcester, MA 01602, USA}
\affiliation{Center for Astrophysics $\vert$ Harvard \& Smithsonian, 60 Garden Street, Cambridge, MA 02138, USA}

\author[0000-0001-7866-2686]{Jihye Hwang}
\affiliation{Korea Astronomy and Space Science Institute, 776 Daedeokdae-ro, Yuseong-gu, Daejeon 34055, Republic of Korea}
\affiliation{University of Science and Technology, Korea, 217 Gajeong-ro, Yuseong-gu, Daejeon 34113, Republic of Korea}

% ...




\begin{abstract}

We present 
BISTRO Survey 850~$\mu$m dust emission polarisation observations of the L1495A-B10 region of the Taurus molecular cloud, 
taken at the JCMT.
We observe a roughly triangular network of dense filaments.  
We detect 9 of the dense starless cores embedded within these filaments in polarisation, finding that the plane-of-sky 
orientation of the core-scale magnetic field lies roughly perpendicular to the filaments in almost all cases.
We also find that the large-scale magnetic field orientation measured by Planck
is not correlated with any of the core or filament structures, except
in the case of the lowest-density core.
We propose a scenario for early prestellar evolution that is both
an extension to, and consistent with, previous models, introducing an
additional evolutionary transitional stage between field-dominated and matter-dominated
evolution, 
observed here for the first time.
In this scenario, the cloud collapses first to a sheet-like structure.
Uniquely, we appear to be seeing this sheet almost face-on.
The sheet fragments into filaments, which in turn form cores. However,
the material must reach a certain critical
density before the 
evolution changes from being field-dominated to being matter-dominated.
We measure the sheet surface density and the magnetic field strength at that transition
for the first time and show consistency with an analytical prediction that had previously gone
untested for over 50 years \citep{1965QJRAS...6..265M}.

\end{abstract}


\section{Introduction} 
\label{sec:intro}

\begin{figure*}
	\centering
    \includegraphics[scale=0.50,angle=0]{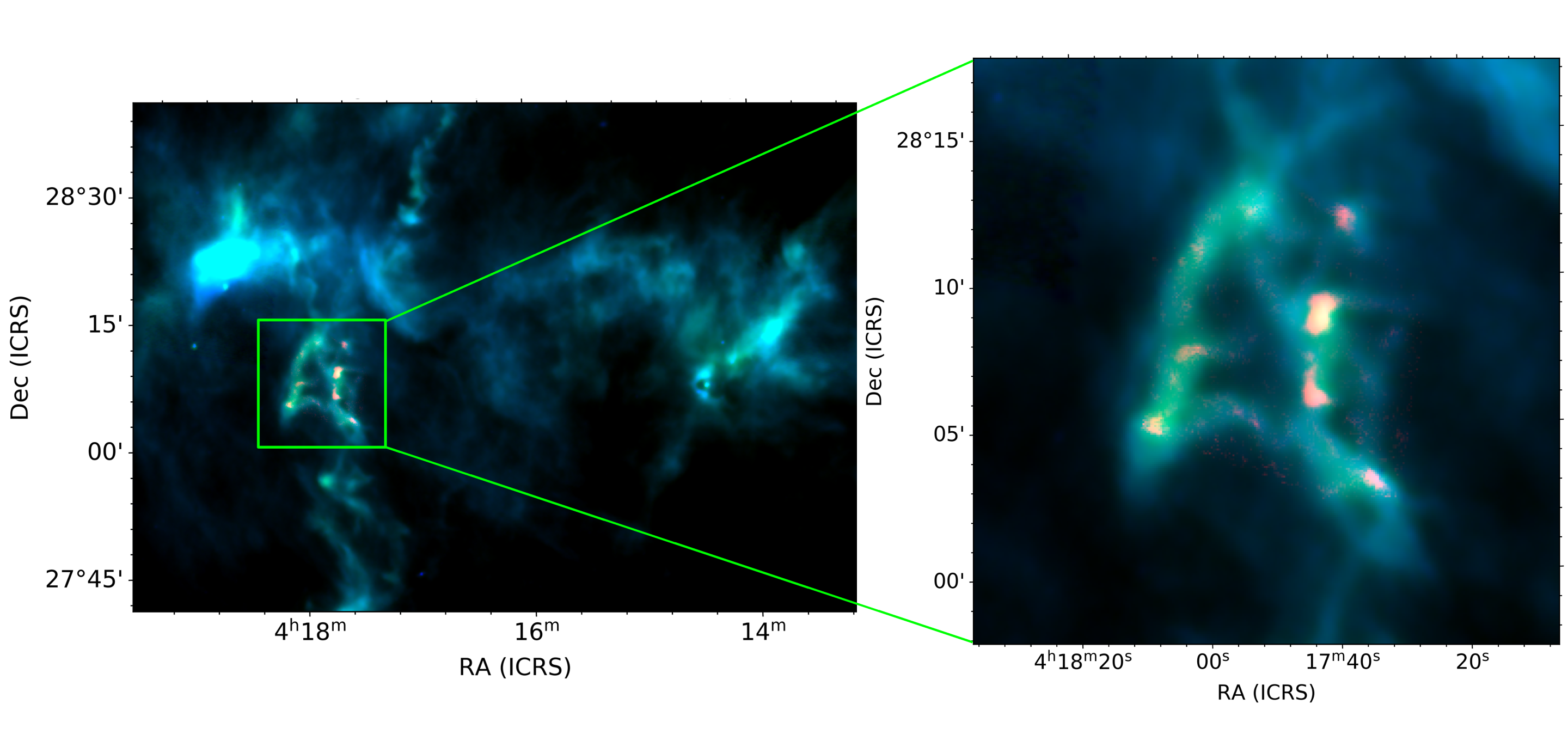}
	\caption{Left: False-colour red-green-blue (RGB) image of the L1495A region taken 
	from the SCUBA2 and \textit{Herschel}-SPIRE observations of 
\citet{2016MNRAS.463.1008W}. Red shows the SCUBA2 850~$\mu$m 
	observations and green and blue show the 500~$\mu$m and 250~$\mu$m 
	\textit{Herschel}-SPIRE observations respectively. 
	This region lies at the head of the L1495 filament, which can be seen heading south from
	the lower left part of the image.
	Right: Zoomed in region of the 
	green square on the left, showing the region mapped in this paper. The RGB channels 
	are the same as in the left panel.}
	\label{fig:fig1}
\end{figure*}

The L1495 region of the Taurus molecular cloud appears as an obscuring dark cloud
on optical images \citep{1907ApJ....25..218B,lynds1962}, with a linear, or
filamentary, structure, coming to a head at a small globule that is
referred to as L1495A \citep{1989ApJS...71...89B,Lee2001}, or sometimes
as B10 \citep[e.g.][]{2018A&A...617A..27P}.

\citet{2016MNRAS.459..342M}
studied the L1495 filament and found it to
be consistent with the filamentary star formation model \citep{Andre2014} favoured by \textit{Herschel} observations of star-forming regions
\citep{2010A&A...518L.102A}. \citet{Lee2001} found evidence for asymmetric
line profiles in the southern part of L1495A, apparently indicating
collapse or contraction. However, this is close to the Herbig Ae/Be
star, V892 Tau, which is clearly affecting at least the southern part
of L1495A, so interpreting asymmetric line profiles is made more
complex in this area. 

\citet{2016MNRAS.463.1008W} presented a comparison of SCUBA2 and \textit{Herschel} observations 
of the L1495A cloud -- see Figure~1. They found that SCUBA2
detects only the highest-surface-brightness sources, principally 
detecting protostellar sources
and starless cores embedded in filaments, while \textit{Herschel} is sensitive to most of the cloud
structure, including extended low-surface-brightness emission. 
For similar temperature
cores, such as those seen in L1495A, the surface brightnesses of the cores are
determined by their column densities, with the highest-column-density cores being detected by
SCUBA2. For roughly spherical geometries, column density can be related to volume density,
and so SCUBA2 selects the densest cores from a population at a given temperature -- thus, the cores in Figure~1 appear red.
Hence, the polarimeter POL2 on SCUBA2 is most sensitive to the polarization in the
cores of L1495A, rather than the filaments.

In this paper we present POL2 data of L1495A as part of the BISTRO survey. In a previous paper from 
the BISTRO survey \citep{2021ApJ...912L..27E} we presented data from the B213
part of the L1495 
filament, which lies to the southeast of this region. 
In both cases we concentrate on the polarisation
of the cores. However, whereas the cores in the previous B213 paper mostly contained YSOs, the cores
studied in this paper are all starless.

\begin{center}
\begin{table*}
    \mycenter
	\caption{Core characteristics}
    \scriptsize
    \begin{tabular}{ccccccccccc}\hline

    Core &RA &Dec. &FWHM$^{b}$  &$\theta_{core}^{b,c}$ & $\theta_{fil}^{c,d}$ &Temp.$^{b}$ &N($H_{2}$)$^{e}$ &n($H_{2}$) &A$_{V}^{f}$ &$\theta_{pol}^{c}$ 

\\
    no.$^{a}$ &(J2000) &(J2000) &($\arcsec \times \arcsec$) &($\degree$) &($\degree$) &(K) &($\times 10^{21} cm^{-2}$) &($\times 10^{5} cm^{-3}$) &(mag) &($\degree$) 

\\ \hline
    1 (2) &4:17:42.10 &+28:08:44.4 & 54.6 $\times$ 21.4 & 167 &26 &10.1 $\pm$ 0.2 &19.1 $\pm$ 7.6 &2.0  $\pm$ 0.8 &17 $\pm$ 7 
    &4 $\pm$ 2 

\\
    2-N (7) &4:17:43.75 &+28:07:04.6 & 32.0 $\times$ 16.0 & 0 &0 &10.6 $\pm$ 0.2 &14.7 $\pm$ 5.9 &2.4  $\pm$ 1.0 & 13 $\pm$ 5 
    &-3 $\pm$ 4 

\\
    2-S (7) &4:17:43.40 &+28:06:04.5 & 32.4 $\times$ 20.7 & 45 &0 &10.6 $\pm$ 0.2 &15.7 $\pm$ 6.3 &2.1  $\pm$ 0.8 & 14 $\pm$ 6 
    &-46 $\pm$ 3 

\\
    3 (12) &4:17:34.58 &+28:03:05.0 & 55.2 $\times$ 20.3 & 53 &37 &12.7 $\pm$ 0.3 &15.3 $\pm$ 6.1 &1.6 $\pm$ 0.6 & 14 $\pm$ 6 
    &68 $\pm$ 11 
 
\\
    4 ($-$) &4:17:53.92 &+28:05:28.3 & 31.4 $\times$ 12.1 & 60 &85 &12.5 $\pm$ 0.1 &9.2 $\pm$ 3.7 &1.7 $\pm$ 0.7 & 8 $\pm$ 3 
    &-72 $\pm$ 8 

\\
    5 (5) &4:18:08.17 &+28:05:10.3 & 39.6 $\times$ 32.0 & 121 &150 &10.1 $\pm$ 0.2 &17.8 $\pm$ 7.1 &1.8 $\pm$ 0.7 & 16 $\pm$ 7 
    &-4 $\pm$ 12 

\\
    6 (11) &4:18:03.08 &+28:07:35.2 & 39.0 $\times$ 20.5 & 126 &170 &9.3 $\pm$ 0.2 &14.0 $\pm$ 5.6 &1.8 $\pm$ 0.7 & 13 $\pm$ 5 
    &18 $\pm$ 5 

\\
    7 (19) &4:18:00.55 &+28:11:08.7 & 45.0 $\times$ 22.2 & 165 &147 &10.7 $\pm$ 0.2 &14.4 $\pm$ 5.8 &1.7 $\pm$ 0.7 & 13 $\pm$ 5 
    &9 $\pm$ 6 
 
\\
    8 (14) &4:17:52.08 &+28:12:31.1 & 51.6 $\times$ 48.7 & 93 &135 (15) &10.9 $\pm$ 0.2 &14.1 $\pm$ 5.6 &1.0 $\pm$ 0.4 & 13 $\pm$ 5 
    &-13 $\pm$ 7

    \\ \hline
	\end{tabular}
	\label{tab:tab1}
    \centering
    \\
	a. Core number in parentheses from \citet{2016MNRAS.463.1008W} \\
	b. Values taken from \citet{2016MNRAS.463.1008W} \\
	c. All angle values are measured east of north \\
	d. We adopt $\pm 10\degree$ for the local filament angle except in cores 4 and 8 (see Section \ref{subsec:f-and-f}) \\
	e. Column density values from Gould Belt Survey \citep{2013A&A...550A..38P} \\
        f. A$_{V}$ calculated using N$_{H}$/A$_{V}\sim2.2\times\,10^{21}$ cm$^{-2}$\,mag$^{-1}$ \citep{2009MNRAS.400.2050G} and N($H_{2}$)$\sim$0.5 N$_{H}$
\end{table*}
\end{center}

\section{Observations and Data Reduction}
\label{sec:obsanddata}

\subsection{SCUBA2-POL2 Observations}
\label{subsec:obs} 

L1495 was observed at 450 and 850~$\mu$m, using SCUBA2-POL2 on the JCMT. 
SCUBA2 only detects Stokes I, whereas POL2 also detects Q and U.
Observations were carried out as part of the BISTRO large program between January 2020 and January 2021 (Project ID: M17BL011). Additional observations as part of BISTRO were further carried out between March and November 2021 (Project ID: M20AL018). There were 20 observations completed, each of $\approx$41 minutes for a total on source time of $\approx$14 hours. The observations were carried out in weather bands 1 and 2 which is equivalent to atmospheric opacity at 225 GHz ($\tau_{225}$) $\le$0.05 
for the first sets of observations
and in the range 0.05--0.08 for the later observations respectively. 

The JCMT has a primary dish diameter of 15~m and a beamsize of 14$\farcs$6 at 850~$\mu$m when approximated with a two-component Gaussian  \citep{2013MNRAS.430.2534D}. 
This corresponds to roughly 0.01~pc at a nominal distance to Taurus of 140~pc \citep{Torres_2009,Schlafly_2014,2020A&A...638A..85R}.
The observations were carried out using a SCUBA2 DAISY mode that is optimized for POL2 observations.
This produces a central 3$\arcmin$ diameter region with uniform coverage and exposure time.
Coverage decreases and noise increases towards the edge of the field of view
\citep{2013MNRAS.430.2513H}. 
This mode has a scan speed of~8~arcsec/sec and a half-wave plate with a rotation frequency of 2~Hz
\citep{2016SPIE.9914E..03F}.

\subsection{Data Reduction}
\label{subsec:data}

The data were reduced using the Submillimetre User Reduction Facility (SMURF) package \citep{2013MNRAS.430.2545C} from the Starlink software \citep{2014ASPC..485..391C}. The SMURF package contains the data reduction routine for SCUBA2-POL2 observations named {\tt\string pol2map} which is used for both 450 and 850~$\mu$m. In this paper we only consider the 850-$\mu$m data.

Initially, the raw bolometer timestreams are separated into separate Stokes \textit{I}, \textit{Q} and \textit{U} timestreams. The command {\it{makemap}} \citep{2013MNRAS.430.2545C} is then called to create an initial Stokes \textit{I} map from the Stokes \textit{I} timestreams.  
The second step of the reduction creates the final Stokes \textit{I}, \textit{Q} and \textit{U} maps
and a polarization half-vector catalog. 
Note that we refer to the polarization measurements as half-vectors because they have
a 180$^\circ$ ambiguity.

We included the parameters {\it{skyloop}} and {\it{mapvar}} in our reduction \citep{2013MNRAS.430.2545C}. {\it{Skyloop}} allows for each observation to be compared to each other at each iteration, as opposed to the previous map-making method where each individual set of observations would iterate individually and then be coadded at the end. Using {\it{skyloop}} allows us to pick up fainter, more extended emission. 

Normally, variances in the map are calculated as linear combinations of the variances of each observation (where variances are calculated by the spread of bolomoter signals which fall within the pixel). However when using {\it{mapvar}}, variances are calculated by the spread across the individual observations. 

The reduction was carried out using a pixel size of 8$\arcsec$. This is a larger pixel size than previous SCUBA2-POL2 reductions which tend to use a pixel size of 4$\arcsec$.
However, with lower flux density starless sources such as those in L1495A, a larger pixel size 
increases the signal-to-noise ratio of the Stokes \textit{I}, \textit{Q} and \textit{U} maps.

The 850-$\mu$m Stokes \textit{I}, \textit{Q} and \textit{U} maps were also multiplied by a Flux Conversion Factor (FCF) of 748 Jy~beam$^{-1}$~pW$^{-1}$ to convert from pW to Jy~beam$^{-1}$ and account for loss of flux from POL2 inserted into the telescope. This value is the standard 495 Jy~beam$^{-1}$~pW$^{-1}$ for reductions using 4$\arcsec$ pixels, multiplied by the standard 1.35 factor from POL2 \citep{2021AJ....162..191M} and then multiplied by a factor of 1.12 to account for the 8$\arcsec$ pixels. This factor was determined from SCUBA2 calibration plots.\footnote{\url{https://www.eaobservatory.org/jcmt/instrumentation/continuum/SCUBA2/calibration}}

To further increase the signal-to-noise ratio (SNR) of our polarization half-vectors, we binned them to a resolution of 12$\arcsec$ as is standard in previous BISTRO work and to roughly match the primary beam size of the JCMT at 850~$\mu$m \citep{2013MNRAS.430.2534D}. The polarization half-vectors are also debiased as described by \citet{1974ApJ...194..249W} to remove statistical bias in regions of low SNR (see Equation \ref{eq:pol}).

The values for the debiased degree of polarization $P$  were calculated from 

\begin{equation}
\label{eq:pol}
P=\frac{1}{I}\sqrt{Q^{2}+U^{2}-\frac{1}{2}(\delta Q^{2}+\delta U^{2})}   \,\, ,
\end{equation}

\noindent
where \textit{I}, \textit{Q}, and \textit{U} are the Stokes parameters, and $\delta Q$, and $\delta U$ are the uncertainties in Stokes \textit{Q} and \textit{U}. The uncertainty $\delta P$ of the degree of polarization was obtained using

\begin{equation}
\delta P = \sqrt{\frac{(Q^2\delta Q^2 + U^2\delta U^2)}{I^2(Q^2+U^2)} + \frac{\delta I^2(Q^2+U^2)}{I^4}}  \,\, ,
\end{equation}

\noindent
with $\delta I$ being the uncertainty in Stokes~\textit{I} (total intensity). 

The polarization position angles $\theta$, measured from north through east on the plane of the sky, were calculated using the relation 

\begin{equation}
{\theta = \frac{1}{2}{\rm tan}^{-1}\frac{U}{Q}} \, .
\label{eq:theta}
\end{equation}

\noindent
The corresponding uncertainties in $\theta$ were calculated using

\begin{equation}
\delta\theta = \frac{1}{2}\frac{\sqrt{Q^2\delta U^2+ U^2\delta Q^2}}{(Q^2+U^2)} \times\frac{180\degree}{\pi}  \,\, .
\label{eq:dtheta}
\end{equation}

\noindent 
We chose SNR cuts of I/$\delta$I $>$ 10 and P/$\delta$P $>$ 2, as is standard for making 
SCUBA2-POL2 polarisation maps. This corresponded to a minimum flux density of around
10~mJy/beam.
The number of individual positions at which we detect polarisation at this level 
varies in each core from a couple up to 20 or more.
Typically, all of the half-vectors in each core were 
seen to be lying roughly parallel to each other, so
we subsequently averaged the measured polarisation half-vectors as a weighted mean
(weighted according to their SNR),
averaged over the FWHM size of each core (see Table~\ref{tab:tab1}),
to make a single polarisation measurement on each core.

The plane-of-sky orientation of the magnetic field is inferred by rotating 
the polarization angles by 90$\degree$, assuming that the polarization is 
caused by elongated dust grains aligned perpendicular to the magnetic 
field \citep[see][and references therein]{ALV2015}. 
This is true for dust grains in the Rayleigh regime \citep{2019MNRAS.488.1211K,2020A&A...634L..15G},
and for observations at 850~$\mu$m this is fulfilled for dust grains 
with sizes up to roughly 100~$\mu$m, which form the majority of those we observe.
This single measurement of the B-field is what is plotted at the position of each core in 
Figure~2 and listed in Table~1 below.

\section{Results} 
\label{sec:result}

\begin{figure}
	\centering
    \includegraphics[scale=0.375,angle=0]{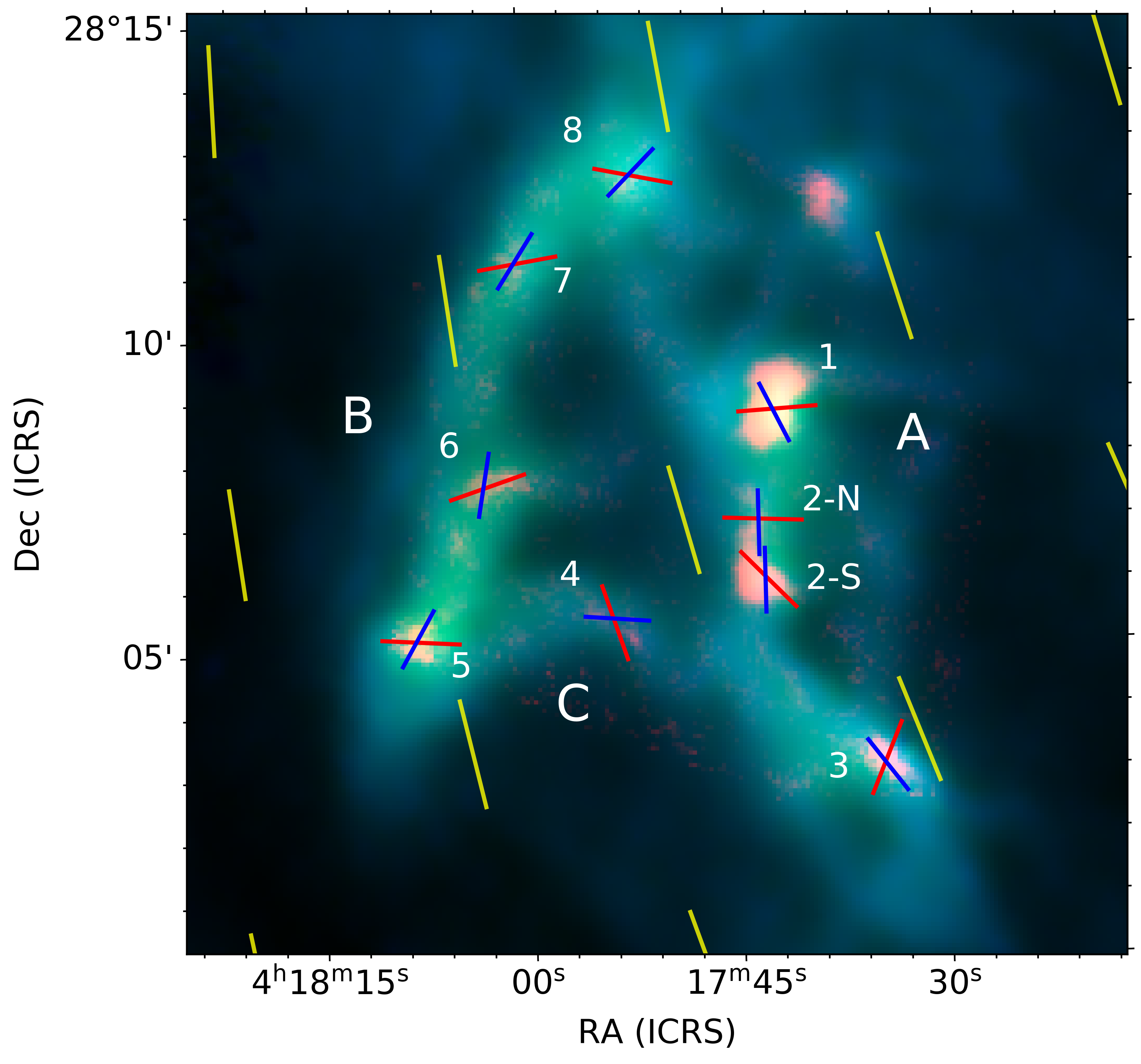}
	\caption{Same background image as Figure \ref{fig:fig1}. 
	The filaments seen in the original SCUBA2 and \textit{Herschel} images are clearly seen and
	are labelled `A', `B' and `C', as described in the text. The cores that we identify in this 
	paper are numbered 1 to 8 as described in the text.
	Superposed on this image are
	rotated polarization half-vectors to demonstrate the orientation of the magnetic field. 
	The red vectors show the mean magnetic field orientation in each core from the POL2 observations.
	The yellow vectors show the orientation of the large-scale magnetic field (over-sampled) from Planck observations
	(there are only about 4 independent Planck beams in this whole field of view). The blue 
	half-vectors 
	show the local filament major axis orientation
	(for core 8 we only show the axis of filament B).
	Each set of vectors in this image has a constant length for clarity.} 
	\label{fig:fig2}
\end{figure}

\subsection{Cores}
\label{subsec:cores}

Figure \ref{fig:fig1} (left) shows a false-colour image (adapted from
\citet{2016MNRAS.463.1008W}), in which green
and blue are taken from the \textit{Herschel}-SPIRE 
images of the region at 500 and 250 $\mu$m respectively, and 
red is the SCUBA2 850-$\mu$m Stokes \textit{I} emission. The much larger-scale B211-213
L1495 filament \citep[e.g.][]{2016MNRAS.459..342M} can be seen 
beginning
at the southern edge of this field and extending southwards out of the field of view. Figure 
\ref{fig:fig1} 
(right) shows a close-up of the triangle of filamentary structures observed by 
\citet{2016MNRAS.463.1008W}. This is the region observed by BISTRO. 
The dense cores show up clearly in the 850-$\mu$m
data as red sources, as mentioned above. These are all starless cores and candidate pre-stellar cores
\citep{2016MNRAS.463.1008W,2019MNRAS.489..962H}.

We here assign the cores numbers as shown in Figure~\ref{fig:fig2}, which has the same background
as Figure~\ref{fig:fig1} (right).
One core where we have detected polarisation was not 
identified as a core by \citet{2016MNRAS.463.1008W}, as it was only detected very 
faintly in the earlier work.
This is labelled on Figure~\ref{fig:fig2} as core 4. 
Hence, if we include core 4 amongst the list of cores detected earlier, we find that
we only detect polarisation at this level on the cores themselves.
We also note that core 2 appears to be split into two cores in the Stokes I image, 
which we here label 2N and 2S. Core 2 was not identified as a double core by \citet{2016MNRAS.463.1008W}.
The two cores 2N and 2S have very different polarisation orientations. 
We therefore have a sample of 9 cores.

Table~\ref{tab:tab1} lists the cores detected in L1495A here in our BISTRO POL2 survey. 
We also state 
the core numbers assigned to them 
by \citet{2016MNRAS.463.1008W}, who fitted elliptical gaussians to 
each of the cores and we list the properties of those
ellipses also in Table~\ref{tab:tab1} (recall that 30$^{\prime\prime}$ is roughly 0.02~pc at a nominal
distance of 140~pc to Taurus).
We note that core 21 from \citet{2016MNRAS.463.1008W} is in our field of view, but no detectable
polarised emission was observed at this position.
For cores 2N, 2S and 4 we calculated the parameters of the elliptical gaussians here.

Table~\ref{tab:tab1} also lists the orientation of the local filament major
axis at the position of each core, which we measure here from Figure~\ref{fig:fig2},
and the core major axis orientation.
Core 8 has two filament angles listed, because it sits at the junction of two
filaments. The value that is in brackets is for the western filament (see also below). 

Table~\ref{tab:tab1} further lists the chief core parameters of temperature as 
measured by \citet{2016MNRAS.463.1008W} and column density which was calculated by 
\citet{2013A&A...550A..38P} using the \textit{Herschel} bands at the resolution of SPIRE 250~$\mu$m. 
We calculate the volume number density in Table~\ref{tab:tab1} using the column density values and 
the 850~$\mu$m core sizes from \citet{2016MNRAS.463.1008W}. The optical extinction values from 
Table~\ref{tab:tab1} are calculated using the column density values. We also list the weighted 
mean of the polarisation position angle in each core from this work measured north through east.

\begin{figure}
	\centering
    \includegraphics[scale=0.3,angle=0]{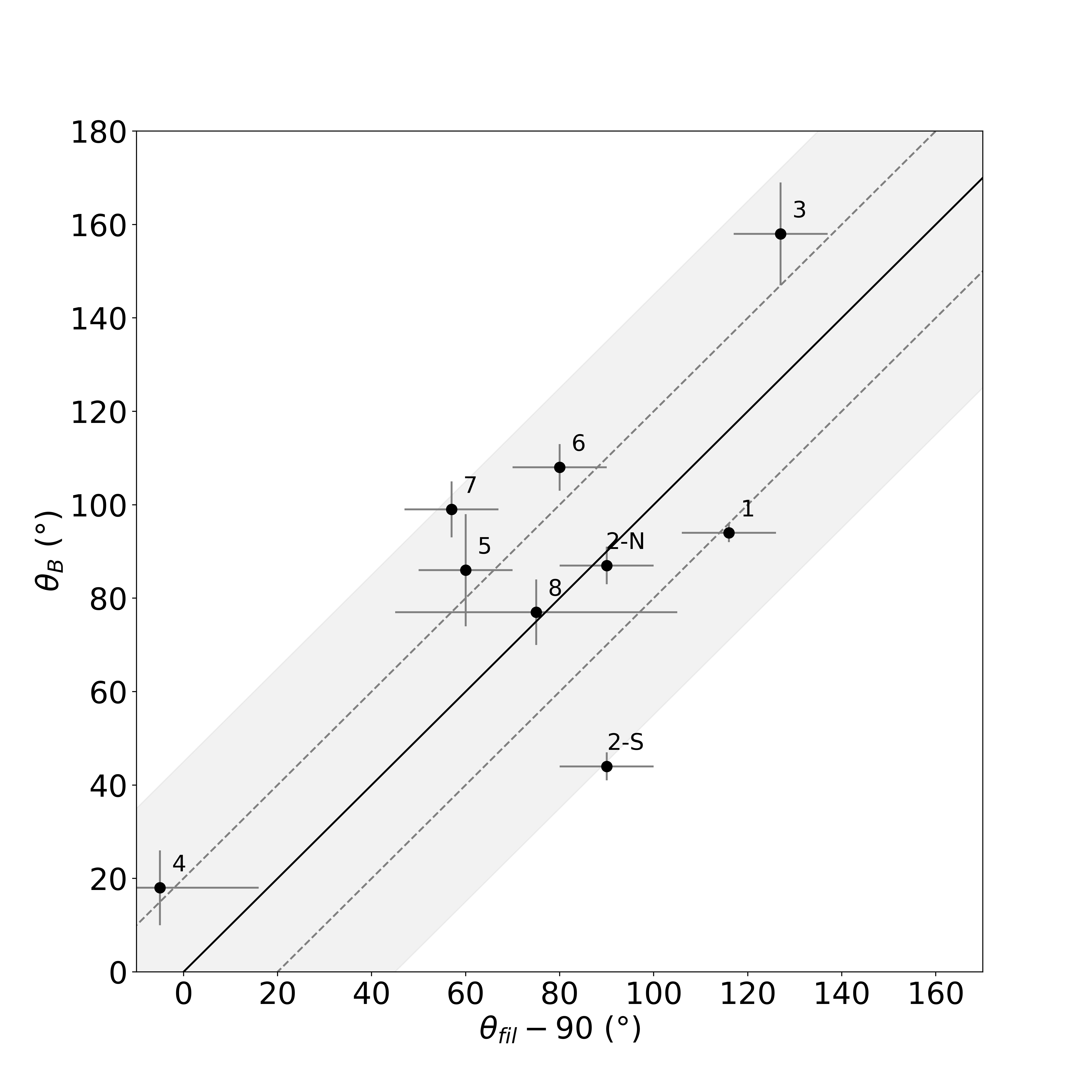}
	\caption{A plot of core magnetic field orientation for each core on the y-axis, versus 90$^o$ minus its local filament major axis angle on the x-axis. The cores are numbered as in Figure~\ref{fig:fig2}. The solid line indicates a one-to-one correlation, which is where the points would be located if the B-field lay exactly orthogonal to the local filament orientation in every case. The two dashed lines represent $\pm$20$^o$, roughly matching our predicted systematic angle error at our chosen signal-to-noise ratio cut-off. 
 The shaded area is $\pm$45$^\circ$.
 See text for details.} 
	\label{fig:fig3}
\end{figure}

Figure \ref{fig:fig2} shows
the SCUBA2-POL2 polarisation half-vectors overlaid in red,
rotated by 90$^o$ to indicate the orientation of the plane-of-sky magnetic field,
within each dense core. As explained above, we refer to them as half-vectors because they 
have a 180$^o$ ambiguity
 (i.e. we don't know which end of the B-field half-vectors 
to put the arrow on). 
Each half-vector
represents the weighted
mean polarisation angle measured for each core respectively, rotated by 90$^o$.

All half-vectors are shown at the same length for clarity.
Also shown on Figure \ref{fig:fig2} in yellow are the half-vectors of the Planck measurements (over-sampled), also
rotated by 90$^o$ to indicate the large-scale plane-of-sky magnetic field orientation. Additionally, 
the blue vectors in each core indicate the local filament major axis as listed in
Table~\ref{tab:tab1}. 





\subsection{Fields and filaments}
\label{subsec:f-and-f}

Examination of Figure~\ref{fig:fig2} shows a number of interesting features. 
The filaments form a roughly triangular pattern on the sky.
We here label the western filament `Filament A', the eastern filament `Filament B'
and the southern filament that runs roughly east-west `Filament C'.
Cores 1, 2 and 3 lie in Filament A, with cores 5, 6 and 7 in Filament B.
Core 4 lies in Filament C.
Core 8 lies at the northern apex of the triangle where Filaments A and B meet.

Figure~\ref{fig:fig2} shows that the magnetic field in cores 1, 2N and 3 lies roughly 
orthogonal to the local filament
(Filament A) long axis.
In core 2S the magnetic field is not orthogonal, but at this point the filament 
turns through 90$^o$,
so it is difficult to uniquely define a filament orientation.
Core 4 has a magnetic field orientation roughly orthogonal 
to the western half of Filament C,
which turns somewhat at the position of core 4. The magnetic field orientations in
cores 5, 6 and 7 also lie roughly orthogonal to their local filament (Filament B) major axis
orientation. Core 7 is slightly further from orthogonal than the other 
two, although we note that here 
also the local filament orientation turns slightly.
The magnetic field of
core 8 is roughly orthogonal to filament B.

Figure \ref{fig:fig3} shows a plot of core magnetic field orientation 
for each core on the y-axis, versus 
90$^o$ minus its local filament major axis angle
on the x-axis, in order to quantify the above discussion.
Any angle that lay between 180$^o$ and 360$^o$ has had 180$^o$ subtracted from it and
any angle that lay between 0$^o$ and $-$180$^o$ has had 180$^o$ added to it due
to the fact that both the magnetic field orientation and the filament orientation are half-vectors,
as discussed above.
The exception to this was core 4, whose error-bar overlaps the origin, so 
we extend the plot to slightly negative numbers to accommodate core 4.

\begin{figure*}
	\centering
    \includegraphics[scale=0.38,angle=0]{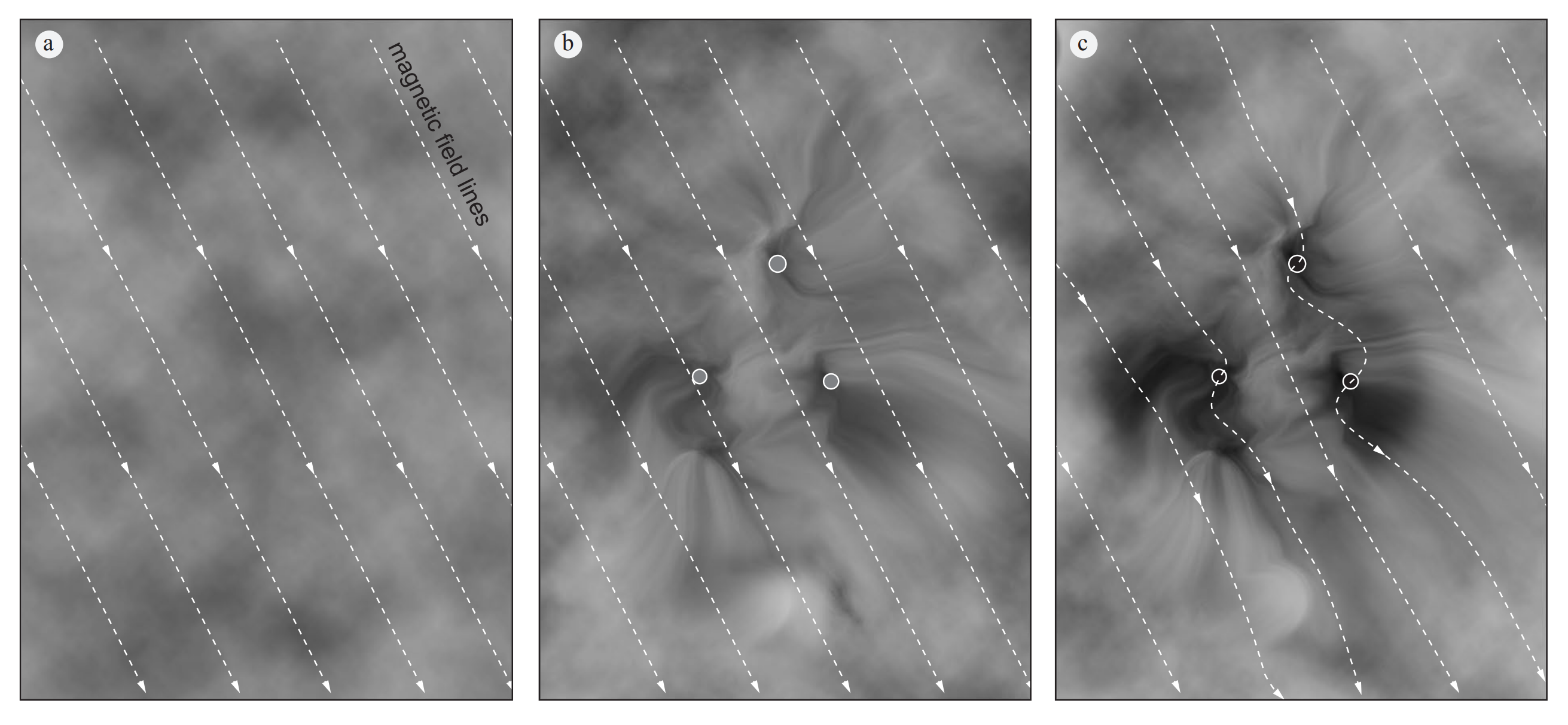}
	\caption{Artist's impression of the three stages of evolution of the filaments and cores in
	L1495A-B10. The grey-scale
 represents the observer's view of column density, 
 as seen in Figure~\ref{fig:fig2}. The dashed lines with arrows represent the magnetic field and circles represent the
 positions of cores.
 The magnetic field lies
	initially mostly along the line of sight, with only a small component in the plane 
 of the sky: 
	(a) The magnetically-dominated phase: the initial cloud is threaded by a large-scale, uniform magnetic field,
	largely along the line of sight, as the matter forms a sheet in the plane of the sky;
	(b) The intermediate phase: the sheet starts to fragment and form filaments, in which 
	low-mass core formation starts;
	(c) The matter-dominated phase: once the matter of the fragmenting structures crosses a 
 critical surface density, the evolution
	becomes matter dominated and the
	subsequent evolution of the matter affects
	the magnetic field to the extent that the local field is
	turned to lie perpendicular to the long axis of the local filament orientation at the position of each core.} 
	\label{fig:fig4}
\end{figure*}

We took the error-bar in filament orientation to be $\pm$10$^o$ as an
indication of how accurately we could measure this orientation. For core 4 where the 
filament 
is curving we took the filament orientation to be the tangent to the curve and the error-bar to be
the amount of curvature.
For core 8, which lies at the junction
of two filaments we took the local filament orientation to be 
that of filament B, which is denser. 
The error-bar used in the B-field orientation in each case
is the standard deviation of the angles of 
the weighted mean half-vectors shown in Figure~\ref{fig:fig2}
and listed in Table~\ref{tab:tab1}.

If all B-field orientations lay exactly orthogonal to their respective filaments, then
this plot would show a one-to-one correlation. This is shown by the solid line on
Figure~\ref{fig:fig3}. We also plot two dashed lines, at $+$20$^o$ and $-$20$^o$ from orthogonal.
This is the typical systematic
error that we would expect for a 2-$\sigma$ detection of
polarization \citep{1993A&A...274..968N}, the value we chose for our cut, as described above.
The shaded area represents $\pm$45$^\circ$.

It can be seen that there is a good degree of correlation in this plot,
with most of the points consistent (including error bars) with lying between, or very close to, the
dashed lines, and all lying within the shaded area.
Hence we conclude that the magnetic fields that we have measured in these cores
generally tend to lie closer to
orthogonal than parallel
to the local filament orientation in which the core is embedded.
The core that lies furthest from the correlation line is 2S. This is one of 
the most dense cores
and may have been affected by the proximity of core 2N or the effect of the filament
changing orientation from the south to the north of core 2S. 

A similar plot to Figure~\ref{fig:fig3} looking for a correlation between
B-field orientation and
core major axis shows no correlation
whatsoever. Likewise a plot of our measured small-scale B-field position angles shows no
correlation with the large-scale B-field orientation measured by Planck, as shown in
Figure~\ref{fig:fig2}. 
It can be seen that the local field that we have measured in the starless cores within
the filaments has totally dissociated from the large-scale field orientation seen by Planck,
and there is no correlation between them.

The exception to this is core 4, whose small-scale B-field that we have measured lies
almost exactly parallel to the large-scale B-field orientation measured by Planck -- see
Figure~\ref{fig:fig2}. This may be a coincidence, because the B-field in core 4 also lies roughly
perpendicular to its host filament. However,
we note that core 4 and the filament in which it sits
have the lowest column density of any of our cores by a factor of 1.5--2
(see Table~\ref{tab:tab1}), calculated 
using standard conversion factors \citep{2009MNRAS.400.2050G,2016MNRAS.459..342M}.
Thus it may be the youngest core.

\section{Discussion: An evolutionary scenario}

All of the foregoing leads us to propose an evolutionary scenario for this cloud that is 
illustrated graphically in Figure~\ref{fig:fig4}. This is an
artist's impression of the three stages of evolution of the filaments and cores in
	L1495A-B10. The grey-scale
 represents the observer's view of column density, 
 as seen in Figure~\ref{fig:fig2}. The dashed lines with arrows represent the magnetic field and circles represent the
 positions of cores.

We split this evolution into three distinct phases, 
illustrated by the three panels in Figure~\ref{fig:fig4}. We label these phases the magnetically-dominated 
phase (4a), the intermediate phase (4b) and the matter-dominated phase (4c).
Figure~\ref{fig:fig4}
is an attempt to reproduce the image seen in Figure~\ref{fig:fig2} at different times during this 
cloud's evolution.

\subsection{The magnetically-dominated phase}
\label{subsec:magdom}

This phase is illustrated in Figure~\ref{fig:fig4}(a) and is predicted by theory to be the earliest phase of 
evolution from an interstellar cloud to the formation of a star. Flux-freezing 
in this phase predicts that the magnetic
field should have a significant effect on any structure that is formed.
This phase is also commonly referred to as being
magnetically sub-critical \citep{2004Ap&SS.292..225C}.

Hence we hypothesise a scenario to explain why, in our observations herein of L1495A-B10
on the plane of the sky, the filament orientations 
appear totally unrelated to one another and to the orientation of the large-scale Planck field.
We hypothesise that the B-field lies mostly along the line-of-sight orientation, 
with a small component in the plane of sky. 
Note that
this can be explained, because,
for a small range of viewing orientations around the B-field axis, there will be both
a clearly visible filament, and a significant component of the field apparently
approximately perpendicular to the
filament. 

If the large-scale field does indeed lie close to the line of sight to the cloud, then the
percentage polarisation seen in the Planck data would be lower at this point than elsewhere in Taurus.
We looked at the Planck data (Planck XIX) and found that this is in fact is the case. L1495A-B10 does
indeed lie in a `polarisation hole' (a minimum of percentage polarisation) relative to other nearby
regions of Taurus. Hence, our hypothesis to explain our data is also consistent with the Planck data,
although this does not constitute proof, since polarisation holes can also be caused by other mechanisms.

In this magnetically-dominated phase,
the material moves mostly along the line of
sight, parallel to the B-field, both towards and away from the observer (see Figure~\ref{fig:fig4}a).
These counter-flows towards and away from us, that feed matter into a denser
layer, are assumed to be oblique, and therefore produce a mildly
sheared layer, as one would expect for the general case. Since we are assuming ideal MHD the flow is mostly aligned with the magnetic field (the path of least resistance). 
Any motion of the matter perpendicular to the field lines has to drag the field with it.

Consequently, the material that we see lies mostly in a sheet-like structure close to 
the plane of the sky -- i.e. we are seeing the sheet almost face-on. 

\begin{figure}
	\centering
    \includegraphics[scale=0.15,angle=0]{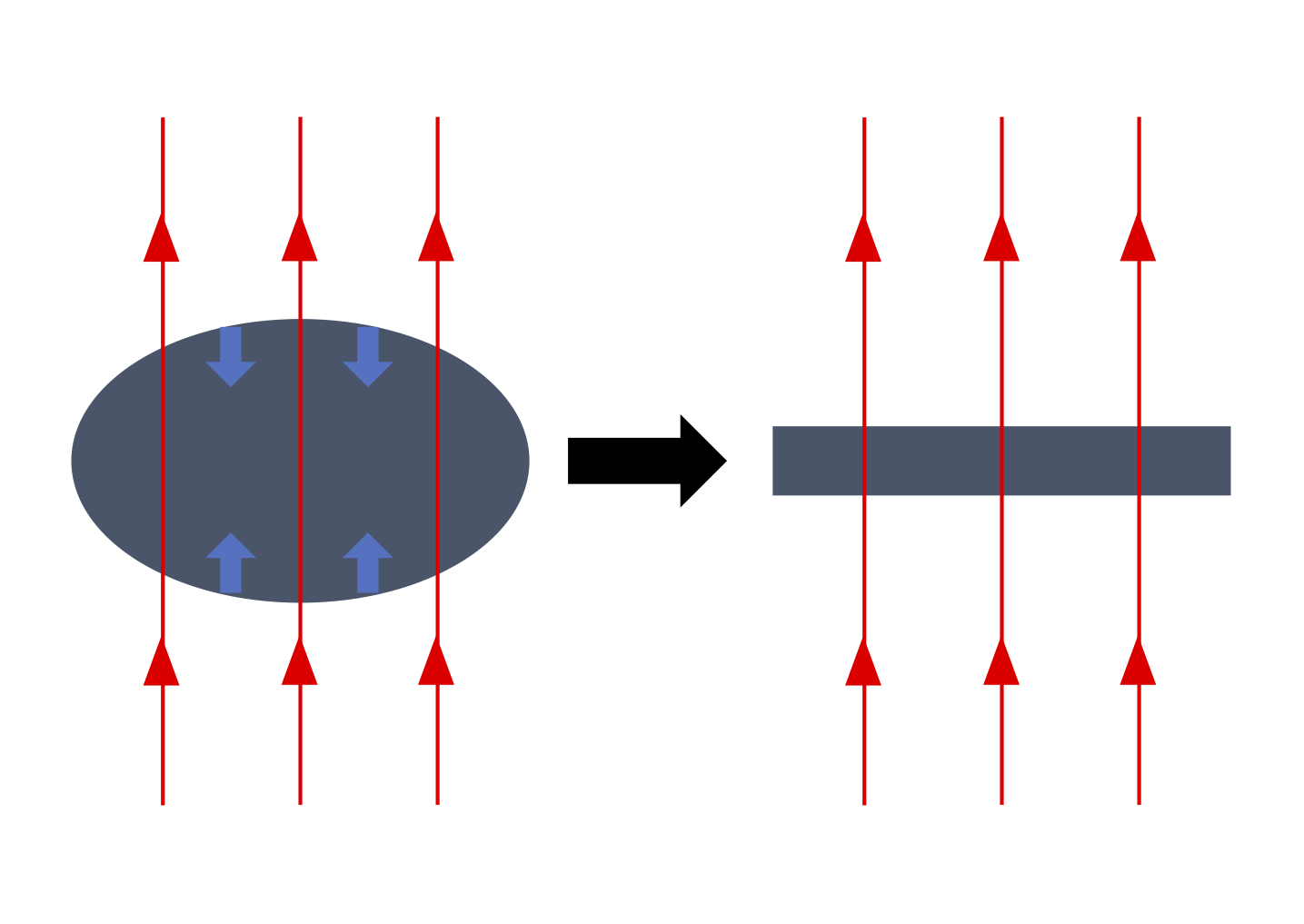}
	\caption{Cartoon illustrating the view roughly orthogonal to the plane-of-sky
views in Figures~\ref{fig:fig2} and \ref{fig:fig4}.
Here the magnetic field (red arrows) is orthogonal to the viewing angle. The blue arrows are indicating the direction the matter is flowing.
The observer is either looking upwards from the bottom of this image or
downwards from the top. The transition from left to right is indicating the
evolution occurring during the phase illustrated in Figure~\ref{fig:fig4}(a),
but here seen edge-on. 
See text for discussion.} 
	\label{fig:fig5}
\end{figure}

We illustrate the evolution occurring during this phase in Figure~\ref{fig:fig5}.
This cartoon illustrates the view roughly orthogonal to the plane-of-sky
views in Figures~\ref{fig:fig2} and \ref{fig:fig4}.
Here the magnetic field (in red) is orthogonal to the viewing angle.
The observer is either looking upwards from the bottom of this image or
downwards from the top.

On the left of Figure~\ref{fig:fig5} we see the initial cloud threaded by a magnetic field.
The blue arrows on the cloud edges indicate the principal directions of collapse along
the magnetic field orientation.
On the right we see the cloud at the end of the magnetically-dominated phase.
The cloud has now collapsed to a sheet-like structure, here seen edge-on, 
orthogonal to the orientation of the magnetic field.

\subsection{The intermediate phase}
\label{subsec:intermediate}

Once the structures are sufficiently dense to cross a certain critical surface density threshold,
the gas motions of the cloud start to transition from magnetically-dominated to matter-dominated.
The sheet fragments under self-gravity and filaments form 
in a fairly complex web-like manner,
all orthogonal to the large-scale magnetic field  (see Figure~\ref{fig:fig4}b).
The convergent velocity field producing each filament
(i.e. an approximately linear over-density within the layer) requires a kink in the
magnetic field, so that
material can slide down the field lines into the filament.

Simulations suggest that the transition to matter-dominated dynamics is associated with a transition from 
the magnetic field being preferentially parallel to density structures at low column densities to 
preferentially perpendicular at high column densities \citep{soler2013}.  This transition appears to require 
converging, likely super-Alfv\'enic, gas flows \citep{soler2017}, and is likely associated with, and 
potentially a signature of, the transition of the region to gravitational instability \citep{chen2016}.
Although the precise mechanism by which (and gas density at which) this transition occurs varies between 
models (see, e.g. \citealt{pattle2022} for a recent review), a large-scale magnetic field strong enough to 
impose a preferred gas flow direction parallel to the field at low densities is typically required 
\citep{soler2017}.

Figure~\ref{fig:fig4}(b) is intended to show the cloud 
in the process of crossing from the magnetically-dominated phase
to the matter-dominated phase. Clearly, this change is not instantaneous and takes a 
finite length of time. The period of the transition is what we here refer to as the intermediate 
phase seen in Figure~\ref{fig:fig4}(b).
We here hypothesise that core 4 may still be in this phase, with its magnetic field
orientation still parallel to that of the initial large-scale field.

\subsection{The matter-dominated phase}
\label{subsec:matterdom}

We would expect
the filaments starting to form low density cores would cause the realignment of 
the magnetic field 
orientations within them (see change from Figure~\ref{fig:fig4}b to c).
This phase is sometimes referred to as being magnetically super-critical \citep{2004Ap&SS.292..225C},
or matter-dominated (our chosen term).

This is now the scenario depicted in Figure~\ref{fig:fig4}(c), which is 
intended to show the picture largely seen in the observations across the 
majority of Figure~\ref{fig:fig2}.

Thus we see that the filaments influence the local small-scale magnetic field orientation, 
turning it to tend to lie apparently roughly orthogonal to the long axis of the
local filament (see Figure~\ref{fig:fig4}c).
Thereafter, the filaments start to form low density cores.
All cores in L1495A-B10 other than core 4 are here hypothesised to be in this phase.

The cores and filaments then start to increase in density due to accretion
of surrounding material. However,
the cores are still at an early 
stage in the star-forming process, since none of them contains a compact protostellar object yet
\citep{2016MNRAS.463.1008W}. Similarly, the filaments are probably also in a fairly
early stage of development.
This is a slight variation on a picture that has previously been proposed -- e.g.
\citet{Andre2014,2019ApJ...871..134S} -- in that it adds an extra transitional
evolutionary phase to the picture that we have here observed for the first time. 

Simulations suggest that a transition in magnetic field orientation within 
molecular clouds from lying parallel to perpendicular to density structure 
occurs only in the case where magnetic fields are dynamically important on 
large scales \citep[e.g.][]{soler2013,seifried2020}.
\citet{planckXXXV} find a significantly sub-critical mass-to-flux ratio of 
$\sim 0.2-0.4$ on large scales in Taurus, while \citet{soler2019} find 
from \textit{Planck} observations that in the L1495/B213 filament (just 
to the south of the region studied here), a transition from preferentially 
parallel to preferentially perpendicular occurs at 
$N_{\rm H}\sim 10^{21.5}\,$cm$^{-2}$.  The balance of evidence suggests 
that the cloud is still magnetically-dominated on large scales.

Note that our scenario for the evolution of L1495A-B10 is consistent with BISTRO 
observations of other nearby filaments with embedded starless cores
(except that the field is not along the line of sight),
including Perseus NGC 1333 \citep{doi2020}, Ophiuchus L1689 \citep{pattle2021} 
and Serpens Main \citep{kwon2022}. All of 
these clouds have magnetic field
orientations consistent with being perpendicular to the 
local filament orientation, and apparently agnostic to the large-scale magnetic 
field orientation in the region.

\subsection{Comparison with theory}

The magnetically- to matter-dominated transition
critical volume density 
is not well-defined. Different theoretical simulations produce transitions at
volume number densities in 
the range $10^{2}-10^{6}\,$cm$^{-3}$, with no
apparent dependence on Alfv\'en Mach number 
\citep{pattle2022}.  We thus need observational constraints on the transition 
critical surface density in a range of environments.

However, calculations of the critical column density at which this transition 
occurs date back over 50 years
\citep[e.g.][-- see his equation 85]{1965QJRAS...6..265M}, 
in which a relatively idealised case of a uniform magnetic
field threads a uniform density cloud \citep[see also][equation 4.82]{2011isf..book.....W}.
Figure~\ref{fig:fig5} shows a simplified cartoon of the picture we are
hypothesising of a cloud collapsing to a sheet, followed by the fragmentation of the
sheet to form filaments and cores (Figure~\ref{fig:fig4}) above a certain sheet critical surface 
density, which
we are referring to as the surface density of the transition from the
magnetically- to the matter-dominated phase.

The above-cited equations lead to a prediction of the relation between the 
critical surface density, $\Sigma_c$, and the magnetic field strength, $B$, at which
the transition occurs, of the form 

\begin{equation}
\Sigma_c \, = \, (5/G)^{1/2} \, (B/3\pi) \,
\label{eq:1}
\end{equation}

\noindent
which yields

\begin{equation}
[N({\rm H_2})/{\rm cm}^{-2}] \, \simeq \, 2 \times 10^{20} \, \times \, [B/\mu G] \, .
\label{eq:2}
\end{equation}

\noindent
This equation contains 
uncertainties of a small numerical factor of order unity \citep[e.g.][]{1978PASJ...30..671N}
but is nevertheless a
useful theoretical order-of-magnitude prediction.

Our hypothesised scenario outlined in Sections~\ref{subsec:magdom}-\ref{subsec:matterdom} above, together with our
interpretation of the data in Section~\ref{subsec:f-and-f} that core~4 may be the youngest core 
and still undergoing the change from the magnetically- to the matter-dominated
phase,
would lead us to set the critical column density  N(H$_2$) 
at around
9.2 $\times$ 10$^{21}$ cm$^{-2}$.
Inserting this value into equation~(\ref{eq:2})
above predicts an order-of-magnitude
magnetic field strength in L1495A-B10 of $\sim$46$\mu$G.
If core 4 has passed the critical point of the intermediate phase, then this B-field value is an upper limit.

This compares to a literature value of 25--77~$\mu$G from optical and NIR observations
\citep{2011ApJ...741...21C} for L1495A-B10.
The equivalent Planck value is 13--32~$\mu$G \citep{planckXXXV} for the large-scale field around
the Taurus region as a whole.

We performed an approximate Davis-Chandrasekhar-Fermi \citep[][DCF]{Davis1951,CF1953} analysis
in core 1, where we have sufficient half-vectors to make this statistical analysis.
For this DCF analysis \citep[e.g.][]{pattle2021}, we used $\Delta V_{NT}$=0.206\,km\,s$^{-1}$ which 
is found from NH$_{3}$ velocity dispersion observations \citep{seo2015} and then removed the thermal component assuming 
a temperature of $\sim$10\,K. We calculated a dispersion in the magnetic field position angle of 11.8$^{o}$ and used the n(H$_{2}$) value of core 1 from Table~\ref{tab:tab1}. We obtained an upper limit of $\approx$70~$\mu$G.

To summarise all of the above: we have hypothesised that L1495A-B10 is evolving from 
the magnetically-dominated phase
to the matter-dominated phase, as the face-on sheet that we are observing starts to fragment 
into filaments and cores.
We have further hypothesised that our observed core 4 may be exactly in that transition phase.
We have used theoretical arguments to predict what field strength would correspond to this transition
at the column densities we observe.
Our theoretical prediction for the field strength required is $\sim$46$\mu$G.
We have estimated the field strength and, along with literature values, found a
range of observed field strengths of 13--32, 25--77 and $\leq$70 $\mu$G.

Hence we see that the whole picture is mutually self-consistent. The theory predicts a B-field
strength that is right in the range of measured values.
Hence the theoretical prediction of \citet{1965QJRAS...6..265M}
is consistent with our observations, and we have managed to fully test this
prediction for the first time in over 50 years.

\subsection{Subsequent evolution}

Using the ratio N$_{H}$/A$_{V}\sim2.2\times\,10^{21}$\,cm$^{-2}$\,mag$^{-1}$ \citep{2009MNRAS.400.2050G}, the corresponding value of A$_{V}$ for core~4 at the column density discussed above
is A$_V\sim 8$ (assuming N($H_{2}$)$\sim$0.5 N$_{H}$).
This is consistent with the star formation threshold 
at which molecular clouds can begin to form stars at 
A$_{V}\sim$7--8, discussed by, e.g.
\citet{2004ApJ...611L..45J,2015A&A...584A..91K,2016MNRAS.459..342M},
among others, and with the observations collated by
\citet{soler2019}, given that it has not yet begun to form stars.

The subsequent behaviour of magnetic fields in regions with embedded 
young stellar objects (YSOs)
appears to be more complex.  Notably, in Taurus B213, a more evolved region located 
elsewhere in the filament of which L1495A-B10 forms the head, the average magnetic field orientation
within the embedded dense cores varies significantly, with no clear correlation with either 
filament or outflow orientation \citep{eswaraiah2021}.  More generally, YSO outflows appear
to have the potential to reshape magnetic fields in their vicinities
\citep[e.g.][]{yen2021,lyo2021,pattle2022a}, suggesting that we cannot infer the magnetic 
initial conditions for star formation in environments with embedded YSOs.

Therefore,
as a region without embedded YSOs, L1495A-B10 represents a more pristine environment in which to study
the behaviour of magnetic fields in the early stages of filament fragmentation and core formation.

\section{Summary}
\label{sec:summary}

In this paper
we have presented the first polarization data from 
BISTRO of the dark cloud L1495A-B10 in the Taurus molecular cloud, 
at a wavelength of 850~$\mu$m.
The cloud contains
a complex pattern of small-scale
filaments, in a roughly triangular morphology, surrounded by a web
of lower density filaments. 
The main filaments contain cores of varying densities. We detect
polarization from 9 of the cores, all of which have previously
been shown to be starless. 

We calculate
the mean plane-of-sky orientation of the small-scale magnetic field in each of the cores
and compare this to the large-scale field orientation measured by Planck and to the orientation 
of the filaments and long axes of the cores. 
There is no correlation between our measurements and those of Planck other than in
the lowest density case, core 4, as expected for the `youngest'
core.
There is also no correlation at all with the core orientation. 

However, we find a correlation between our small-scale 
field measurements and filament orientation, 
in which the field tends to lie orthogonal to the
long axes of the filaments in all cases except for a core that lies near to a bend in 
its local filament.

We hypothesise a scenario to explain our data, which is an extension of the
previous magnetic filamentary model, adding an additional
transitional evolutionary phase to the
model, that we here observe for the first time. 
We suggest we are seeing a cloud which was originally magnetically-dominated in the process of transition to being matter-dominated in its denser regions, with core 4 defining the transition point, 
since the local field in core 4
is still aligned with the large-scale field, whereas in the others it tends more towards lying orthogonal 
to the local filament.

We measure the sheet surface density and the magnetic field strength of that transitionary phase
for the first time and show consistency with the original analytical prediction that has gone
untested for over 50 years \citep{1965QJRAS...6..265M}.

\bigskip
\bigskip

D.W.T. acknowledges funding support from the UK STFC through grant number ST/R000786/1.
J.K. acknowledges funding from the Moses Holden Scholarship for his PhD.  K.P is a Royal Society University Research Fellow, supported by grant number URF\textbackslash R1\textbackslash211322 The JCMT is operated by the East Asian Observatory on behalf of National Astronomical Observatory of Japan; 
the UK STFC under the auspices of grant number ST/N005856/1;
Academia Sinica Institute of Astronomy and Astrophysics; the Korea Astronomy and Space Science Institute; the Operation, Maintenance and Upgrading Fund for Astronomical Telescopes and Facility Instruments, budgeted from the Ministry of Finance of China. SCUBA2 and POL2 were built through grants from the Canada Foundation for Innovation. This research used the facilities of the Canadian Astronomy Data Centre operated by the National Research Council of Canada with the support of the Canadian Space Agency. The data taken in this paper were observed under the project code M20AL018. The Starlink software \citep{2014ASPC..485..391C} is currently supported by the East Asian Observatory.  The authors wish to recognize and acknowledge the very significant cultural role and reverence that the summit of Maunakea has always had within the indigenous Hawaiian community. We are most fortunate to have the opportunity to conduct observations from this mountain.
M.T. is supported by JSPS KAKENHI grant Nos.18H05442, 15H02063, and 22000005. J.K.is supported by JSPS KAKENHI grant No.19K14775.
F.P. acknowledges support from the Spanish State Research Agency (AEI) under grant number
PID2019-105552RB-C43.
F.K. is supported by the Spanish program Unidad de Excelencia Mar\'ia de Maeztu CEX2020-001058-M, financed by MCIN/AEI/10.13039/501100011033.
CE acknowledges the financial support from grant RJF/2020/000071 as a part of Ramanujan Fellowship awarded by the Science and Engineering Research Board (SERB), Department of Science and Technology (DST), Government of India.
W.K. was supported by the National Research Foundation of Korea (NRF) grant funded by the Korea government (MSIT) (NRF-2021R1F1A1061794).
The work of M.G.R. is supported by NOIRLab, which is managed by the Association of Universities for Research in Astronomy (AURA) under a cooperative agreement with the National Science Foundation.

\vspace{5mm}
\facilities{JCMT (SCUBA2, POL2)}


\software{Starlink \citep{2014ASPC..485..391C}, Astropy \citep{2013A&A...558A..33A,2018AJ....156..123A}}




\bibliography{citations}{}
\bibliographystyle{aasjournal}

\end{document}